\documentstyle[10pt,epsf,epsfig,dp_delphititle]{dp_delphi}
\makeindex
\def\DpPaperGroup{EP}
\def\DpPaperRef{2000-110}
\def\DpDate{16 August 2000}
\def\DpAuthors{DELPHI Collaboration}
\def\DpSubmit{(Submitted to Phys.Lett.B)}
\def\DpTitle{{Search for the sgoldstino at $\sqrt{s}$ 
from 189 to 202 GeV }}
\def\DpComment{ }
\def\DpEMail{ }
\begin{document}
\makeatletter
\newcount\@tempcntc
\def\@citex[#1]#2{\if@filesw\immediate\write\@auxout{\string\citation{#2}}\fi
  \@tempcnta\z@\@tempcntb\m@ne\def\@citea{}\@cite{\@for\@citeb:=#2\do
    {\@ifundefined
       {b@\@citeb}{\@citeo\@tempcntb\m@ne\@citea\def\@citea{,}{\bf ?}\@warning
       {Citation `\@citeb' on page \thepage \space undefined}}%
    {\setbox\z@\hbox{\global\@tempcntc0\csname b@\@citeb\endcsname\relax}%
     \ifnum\@tempcntc=\z@ \@citeo\@tempcntb\m@ne
       \@citea\def\@citea{,}\hbox{\csname b@\@citeb\endcsname}%
     \else
      \advance\@tempcntb\@ne
      \ifnum\@tempcntb=\@tempcntc
      \else\advance\@tempcntb\m@ne\@citeo
      \@tempcnta\@tempcntc\@tempcntb\@tempcntc\fi\fi}}\@citeo}{#1}}
\def\@citeo{\ifnum\@tempcnta>\@tempcntb\else\@citea\def\@citea{,}%
  \ifnum\@tempcnta=\@tempcntb\the\@tempcnta\else
   {\advance\@tempcnta\@ne\ifnum\@tempcnta=\@tempcntb \else \def\@citea{--}\fi
    \advance\@tempcnta\m@ne\the\@tempcnta\@citea\the\@tempcntb}\fi\fi}
 
\makeatother
\begin{titlepage}
\pagenumbering{roman}
\CERNpreprint{\DpPaperGroup}{\DpPaperRef} 
\date{{\small\DpDate}} 
\title{\DpTitle} 
\address{\DpAuthors} 
\begin{shortabs} 
\noindent
%
\noindent
A search for the supersymmetric partner of the goldstino, the sgoldstino S,
at LEP2 is presented. The production S$\gamma$  followed 
by S decay into two gluons or into two photons was studied at 
189 - 202 GeV LEP centre-of-mass energies. No evidence for the S production 
was found and limits on the S mass corresponding to different
theory parameters are given.

\end{shortabs}
\vfill
\begin{center}
\DpSubmit \ \\ 
\DpComment \ \\
\DpEMail \ \\
\end{center}
\vfill
\clearpage
\headsep 10.0pt
\addtolength{\textheight}{10mm}
\addtolength{\footskip}{-5mm}
\begingroup
%
\newcommand{\DpName}[2]{\hbox{#1$^{\ref{#2}}$},\hfill}
\newcommand{\DpNameTwo}[3]{\hbox{#1$^{\ref{#2},\ref{#3}}$},\hfill}
\newcommand{\DpNameThree}[4]{\hbox{#1$^{\ref{#2},\ref{#3},\ref{#4}}$},\hfill}
\newskip\Bigfill \Bigfill = 0pt plus 1000fill
\newcommand{\DpNameLast}[2]{\hbox{#1$^{\ref{#2}}$}\hspace{\Bigfill}}
%
\footnotesize
\noindent
\DpName{P.Abreu}{LIP}
\DpName{W.Adam}{VIENNA}
\DpName{T.Adye}{RAL}
\DpName{P.Adzic}{DEMOKRITOS}
\DpName{Z.Albrecht}{KARLSRUHE}
\DpName{T.Alderweireld}{AIM}
\DpName{G.D.Alekseev}{JINR}
\DpName{R.Alemany}{VALENCIA}
\DpName{T.Allmendinger}{KARLSRUHE}
\DpName{P.P.Allport}{LIVERPOOL}
% ...
\normalsize
\endgroup
\titlefoot{Department of Physics and Astronomy, Iowa State
     University, Ames IA 50011-3160, USA
    \label{AMES}}
\titlefoot{Physics Department, Univ. Instelling Antwerpen,
     Universiteitsplein 1, B-2610 Antwerpen, Belgium \\
     \indent~~and IIHE, ULB-VUB,
     Pleinlaan 2, B-1050 Brussels, Belgium \\
     \indent~~and Facult\'e des Sciences,
     Univ. de l'Etat Mons, Av. Maistriau 19, B-7000 Mons, Belgium
    \label{AIM}}
\titlefoot{Physics Laboratory, University of Athens, Solonos Str.
     104, GR-10680 Athens, Greece
    \label{ATHENS}}
\titlefoot{Department of Physics, University of Bergen,
     All\'egaten 55, NO-5007 Bergen, Norway
    \label{BERGEN}}
\titlefoot{Dipartimento di Fisica, Universit\`a di Bologna and INFN,
     Via Irnerio 46, IT-40126 Bologna, Italy
    \label{BOLOGNA}}
\titlefoot{Centro Brasileiro de Pesquisas F\'{\i}sicas, rua Xavier Sigaud 150,
     BR-22290 Rio de Janeiro, Brazil \\
     \indent~~and Depto. de F\'{\i}sica, Pont. Univ. Cat\'olica,
     C.P. 38071 BR-22453 Rio de Janeiro, Brazil \\
     \indent~~and Inst. de F\'{\i}sica, Univ. Estadual do Rio de Janeiro,
     rua S\~{a}o Francisco Xavier 524, Rio de Janeiro, Brazil
    \label{BRASIL}}
\titlefoot{Comenius University, Faculty of Mathematics and Physics,
     Mlynska Dolina, SK-84215 Bratislava, Slovakia
    \label{BRATISLAVA}}
\titlefoot{Coll\`ege de France, Lab. de Physique Corpusculaire, IN2P3-CNRS,
     FR-75231 Paris Cedex 05, France
    \label{CDF}}
\titlefoot{CERN, CH-1211 Geneva 23, Switzerland
    \label{CERN}}
\titlefoot{Institut de Recherches Subatomiques, IN2P3 - CNRS/ULP - BP20,
     FR-67037 Strasbourg Cedex, France
    \label{CRN}}
\titlefoot{Now at DESY-Zeuthen, Platanenallee 6, D-15735 Zeuthen, Germany
    \label{DESY}}
\titlefoot{Institute of Nuclear Physics, N.C.S.R. Demokritos,
     P.O. Box 60228, GR-15310 Athens, Greece
    \label{DEMOKRITOS}}
\titlefoot{FZU, Inst. of Phys. of the C.A.S. High Energy Physics Division,
     Na Slovance 2, CZ-180 40, Praha 8, Czech Republic
    \label{FZU}}
\titlefoot{Dipartimento di Fisica, Universit\`a di Genova and INFN,
     Via Dodecaneso 33, IT-16146 Genova, Italy
    \label{GENOVA}}
\titlefoot{Institut des Sciences Nucl\'eaires, IN2P3-CNRS, Universit\'e
     de Grenoble 1, FR-38026 Grenoble Cedex, France
    \label{GRENOBLE}}
\titlefoot{Helsinki Institute of Physics, HIP,
     P.O. Box 9, FI-00014 Helsinki, Finland
    \label{HELSINKI}}
\titlefoot{Joint Institute for Nuclear Research, Dubna, Head Post
     Office, P.O. Box 79, RU-101 000 Moscow, Russian Federation
    \label{JINR}}
\titlefoot{Institut f\"ur Experimentelle Kernphysik,
     Universit\"at Karlsruhe, Postfach 6980, DE-76128 Karlsruhe,
     Germany
    \label{KARLSRUHE}}
\titlefoot{Institute of Nuclear Physics and University of Mining and Metalurgy,
     Ul. Kawiory 26a, PL-30055 Krakow, Poland
    \label{KRAKOW}}
\titlefoot{Universit\'e de Paris-Sud, Lab. de l'Acc\'el\'erateur
     Lin\'eaire, IN2P3-CNRS, B\^{a}t. 200, FR-91405 Orsay Cedex, France
    \label{LAL}}
\titlefoot{School of Physics and Chemistry, University of Lancaster,
     Lancaster LA1 4YB, UK
    \label{LANCASTER}}
\titlefoot{LIP, IST, FCUL - Av. Elias Garcia, 14-$1^{o}$,
     PT-1000 Lisboa Codex, Portugal
    \label{LIP}}
\titlefoot{Department of Physics, University of Liverpool, P.O.
     Box 147, Liverpool L69 3BX, UK
    \label{LIVERPOOL}}
\titlefoot{LPNHE, IN2P3-CNRS, Univ.~Paris VI et VII, Tour 33 (RdC),
     4 place Jussieu, FR-75252 Paris Cedex 05, France
    \label{LPNHE}}
\titlefoot{Department of Physics, University of Lund,
     S\"olvegatan 14, SE-223 63 Lund, Sweden
    \label{LUND}}
\titlefoot{Universit\'e Claude Bernard de Lyon, IPNL, IN2P3-CNRS,
     FR-69622 Villeurbanne Cedex, France
    \label{LYON}}
\titlefoot{Univ. d'Aix - Marseille II - CPP, IN2P3-CNRS,
     FR-13288 Marseille Cedex 09, France
    \label{MARSEILLE}}
\titlefoot{Dipartimento di Fisica, Universit\`a di Milano and INFN-MILANO,
     Via Celoria 16, IT-20133 Milan, Italy
    \label{MILANO}}
\titlefoot{Dipartimento di Fisica, Univ. di Milano-Bicocca and
     INFN-MILANO, Piazza delle Scienze 2, IT-20126 Milan, Italy
    \label{MILANO2}}
\titlefoot{Niels Bohr Institute, Blegdamsvej 17,
     DK-2100 Copenhagen {\O}, Denmark
    \label{NBI}}
\titlefoot{IPNP of MFF, Charles Univ., Areal MFF,
     V Holesovickach 2, CZ-180 00, Praha 8, Czech Republic
    \label{NC}}
\titlefoot{NIKHEF, Postbus 41882, NL-1009 DB
     Amsterdam, The Netherlands
    \label{NIKHEF}}
\titlefoot{National Technical University, Physics Department,
     Zografou Campus, GR-15773 Athens, Greece
    \label{NTU-ATHENS}}
\titlefoot{Physics Department, University of Oslo, Blindern,
     NO-1000 Oslo 3, Norway
    \label{OSLO}}
\titlefoot{Dpto. Fisica, Univ. Oviedo, Avda. Calvo Sotelo
     s/n, ES-33007 Oviedo, Spain
    \label{OVIEDO}}
\titlefoot{Department of Physics, University of Oxford,
     Keble Road, Oxford OX1 3RH, UK
    \label{OXFORD}}
\titlefoot{Dipartimento di Fisica, Universit\`a di Padova and
     INFN, Via Marzolo 8, IT-35131 Padua, Italy
    \label{PADOVA}}
\titlefoot{Rutherford Appleton Laboratory, Chilton, Didcot
     OX11 OQX, UK
    \label{RAL}}
\titlefoot{Dipartimento di Fisica, Universit\`a di Roma II and
     INFN, Tor Vergata, IT-00173 Rome, Italy
    \label{ROMA2}}
\titlefoot{Dipartimento di Fisica, Universit\`a di Roma III and
     INFN, Via della Vasca Navale 84, IT-00146 Rome, Italy
    \label{ROMA3}}
\titlefoot{DAPNIA/Service de Physique des Particules,
     CEA-Saclay, FR-91191 Gif-sur-Yvette Cedex, France
    \label{SACLAY}}
\titlefoot{Instituto de Fisica de Cantabria (CSIC-UC), Avda.
     los Castros s/n, ES-39006 Santander, Spain
    \label{SANTANDER}}
\titlefoot{Dipartimento di Fisica, Universit\`a degli Studi di Roma
     La Sapienza, Piazzale Aldo Moro 2, IT-00185 Rome, Italy
    \label{SAPIENZA}}
\titlefoot{Inst. for High Energy Physics, Serpukov
     P.O. Box 35, Protvino, (Moscow Region), Russian Federation
    \label{SERPUKHOV}}
\titlefoot{J. Stefan Institute, Jamova 39, SI-1000 Ljubljana, Slovenia
     and Laboratory for Astroparticle Physics,\\
     \indent~~Nova Gorica Polytechnic, Kostanjeviska 16a, SI-5000 Nova Gorica, Slovenia, \\
     \indent~~and Department of Physics, University of Ljubljana,
     SI-1000 Ljubljana, Slovenia
    \label{SLOVENIJA}}
\titlefoot{Fysikum, Stockholm University,
     Box 6730, SE-113 85 Stockholm, Sweden
    \label{STOCKHOLM}}
\titlefoot{Dipartimento di Fisica Sperimentale, Universit\`a di
     Torino and INFN, Via P. Giuria 1, IT-10125 Turin, Italy
    \label{TORINO}}
\titlefoot{Dipartimento di Fisica, Universit\`a di Trieste and
     INFN, Via A. Valerio 2, IT-34127 Trieste, Italy \\
     \indent~~and Istituto di Fisica, Universit\`a di Udine,
     IT-33100 Udine, Italy
    \label{TU}}
\titlefoot{Univ. Federal do Rio de Janeiro, C.P. 68528
     Cidade Univ., Ilha do Fund\~ao
     BR-21945-970 Rio de Janeiro, Brazil
    \label{UFRJ}}
\titlefoot{Department of Radiation Sciences, University of
     Uppsala, P.O. Box 535, SE-751 21 Uppsala, Sweden
    \label{UPPSALA}}
\titlefoot{IFIC, Valencia-CSIC, and D.F.A.M.N., U. de Valencia,
     Avda. Dr. Moliner 50, ES-46100 Burjassot (Valencia), Spain
    \label{VALENCIA}}
\titlefoot{Institut f\"ur Hochenergiephysik, \"Osterr. Akad.
     d. Wissensch., Nikolsdorfergasse 18, AT-1050 Vienna, Austria
    \label{VIENNA}}
\titlefoot{Inst. Nuclear Studies and University of Warsaw, Ul.
     Hoza 69, PL-00681 Warsaw, Poland
    \label{WARSZAWA}}
\titlefoot{Fachbereich Physik, University of Wuppertal, Postfach
     100 127, DE-42097 Wuppertal, Germany
    \label{WUPPERTAL}}
\addtolength{\textheight}{-10mm}
\addtolength{\footskip}{5mm}
\clearpage
\headsep 30.0pt
\end{titlepage}
%
\pagenumbering{arabic} 
\setcounter{footnote}{0} %
\large
\newcommand{\rul}{\rule[-3mm]{0mm}{8mm}}
\newcommand{\trul}{\rule[+3mm]{0mm}{1mm}}
\newcommand{\mc}{\multicolumn}
\newcommand{\m}{\mbox}
%
%
%
\newcommand {\grav}    {\rm{\tilde G}}
\newcommand {\GeV}     {\rm{GeV}}
\newcommand {\MeV}     {\rm{MeV}}
\newcommand {\nb}      {\rm{nb}}
\newcommand {\Zzero}   {{\rm Z}^0}
\newcommand {\MZ}      {\rm{M_Z}}
\newcommand {\MW}      {\rm{M_W}}
\newcommand {\GF}      {\rm{G_F}}
\newcommand {\Gm}      {\rm{G_{\mu}}}
\newcommand {\MH}      {\rm{M_H}}
\newcommand {\MT}      {\rm{m_{top}}}
\newcommand {\GZ}      {\Gamma_{\rm Z}}
\newcommand {\Afb}     {\rm{A_{FB}}}
\newcommand {\Afbs}    {\rm{A_{FB}^{s}}}
\newcommand {\sigmaf}  {\sigma_{\rm{F}}}
\newcommand {\sigmab}  {\sigma_{\rm{B}}}
\newcommand {\NF}      {\rm{N_{F}}}
\newcommand {\NB}      {\rm{N_{B}}}
\newcommand {\Nnu}     {\rm{N_{\nu}}}
\newcommand {\RZ}      {\rm{R_Z}}
\newcommand {\rhob}    {\rho_{eff}}
\newcommand {\Gammanz} {\rm{\Gamma_{Z}^{new}}}
\newcommand {\Gammani} {\rm{\Gamma_{inv}^{new}}}
\newcommand {\Gammasz} {\rm{\Gamma_{Z}^{SM}}}
\newcommand {\Gammasi} {\rm{\Gamma_{inv}^{SM}}}
\newcommand {\Gammaxz} {\rm{\Gamma_{Z}^{exp}}}
\newcommand {\Gammaxi} {\rm{\Gamma_{inv}^{exp}}}
\newcommand {\rhoZ}    {\rho_{\rm Z}}
\newcommand {\thw}        {\theta_{\rm W}}
\newcommand {\swsq}       {\sin^2\!\thw}
\newcommand {\swsqmsb}    {\sin^2\!\theta_{\rm W}^{\overline{\rm MS}}}
\newcommand {\swsqbar}    {\sin^2\!\overline{\theta}_{\rm W}}
\newcommand {\cwsqbar}    {\cos^2\!\overline{\theta}_{\rm W}}
\newcommand {\swsqb}      {\sin^2\!\theta^{eff}_{\rm W}}
\newcommand {\ee}         {{e^+e^-}}
\newcommand {\eeX}        {{e^+e^-X}}
\newcommand {\gaga}       {{\gamma\gamma}}
\newcommand {\gagaga}       {{\gamma\gamma(\gamma)}}
\newcommand {\mumu}       {{\mu^+\mu^-}}
\newcommand {\eeg}        {{e^+e^-\gamma}}
\newcommand {\mumug}      {{\mu^+\mu^-\gamma}}
\newcommand {\tautau}     {{\tau^+\tau^-}}
\newcommand {\tautaug}     {{\tau^+\tau^-\gamma}}
\newcommand {\qqb}        {{q\bar{q}}}
\newcommand {\eegg}       {e^+e^-\rightarrow \gamma\gamma}
\newcommand {\eeggg}      {e^+e^-\rightarrow \gamma\gamma(\gamma)}
\newcommand {\eeee}       {e^+e^-\rightarrow e^+e^-}
\newcommand {\eeeeee}     {e^+e^-\rightarrow e^+e^-e^+e^-}
\newcommand {\eeeeg}      {e^+e^-\rightarrow e^+e^-(\gamma)}
\newcommand {\eeeegg}     {e^+e^-\rightarrow e^+e^-\gamma\gamma}
\newcommand {\eeeg}       {e^+e^-\rightarrow (e^+)e^-\gamma}
\newcommand {\eemumu}     {e^+e^-\rightarrow \mu^+\mu^-}
\newcommand {\eetautau}   {e^+e^-\rightarrow \tau^+\tau^-}
\newcommand {\eehad}      {e^+e^-\rightarrow {\rm hadrons}}
\newcommand {\eenng}      {e^+e^-\rightarrow \nu\bar{\nu}\gamma}
\newcommand {\eeGrGr}      {e^+e^-\rightarrow \tilde{G}\tilde{G}}
\newcommand {\eeGrGrg}      {e^+e^-\rightarrow \tilde{G}\tilde{G}\gamma}
\newcommand {\eeSg}      {e^+e^-\rightarrow S\gamma}
\newcommand {\eettg}      {e^+e^-\rightarrow \tau^+\tau^-\gamma}
\newcommand {\eell}       {e^+e^-\rightarrow l^+l^-}
\newcommand {\Zbb}       {\rm Z^0\rightarrow b\bar{b}}
\newcommand {\Ztopig}    {{\rm Z}^0\rightarrow \pi^0\gamma}
\newcommand {\Ztoetag}    {{\rm Z}^0\rightarrow \eta\gamma}
\newcommand {\Ztoomegag}    {{\rm Z}^0\rightarrow \omega\gamma}
\newcommand {\Ztogg}     {{\rm Z}^0\rightarrow \gamma\gamma}
\newcommand {\Ztoee}     {{\rm Z}^0\rightarrow e^+e^-}
\newcommand {\Ztoggg}    {{\rm Z}^0\rightarrow \gamma\gamma\gamma}
\newcommand {\Ztomumu}   {{\rm Z}^0\rightarrow \mu^+\mu^-}
\newcommand {\Ztotautau} {{\rm Z}^0\rightarrow \tau^+\tau^-}
\newcommand {\Ztoll}     {{\rm Z}^0\rightarrow l^+l^-}
\newcommand {\Lamp}       {\Lambda_{+}}
\newcommand {\Lamm}       {\Lambda_{-}}
\newcommand {\Gee}        {\Gamma_{ee}}
\newcommand {\Gpig}       {\Gamma_{\pi^0\gamma}}
\newcommand {\Ggg}        {\Gamma_{\gamma\gamma}}
\newcommand {\Gggg}       {\Gamma_{\gamma\gamma\gamma}}
\newcommand {\Gmumu}      {\Gamma_{\mu\mu}}
\newcommand {\Gtautau}    {\Gamma_{\tau\tau}}
\newcommand {\Ginv}       {\Gamma_{\rm inv}}
\newcommand {\Ghad}       {\Gamma_{\rm had}}
\newcommand {\Gnu}        {\Gamma_{\nu}}
\newcommand {\GnuSM}      {\Gamma_{\nu}^{\rm SM}}
\newcommand {\Gll}        {\Gamma_{l^+l^-}}
\newcommand {\Gff}        {\Gamma_{f\overline{f}}}
\newcommand {\Gtot}       {\Gamma_{\rm tot}}
\newcommand {\al}         {a_l}
\newcommand {\vl}         {v_l}
\newcommand {\af}         {a_f}
\newcommand {\vf}         {v_f}
\newcommand {\ael}        {a_e}
\newcommand {\ve}         {v_e}
\newcommand {\amu}        {a_\mu}
\newcommand {\vmu}        {v_\mu}
\newcommand {\atau}       {a_\tau}
\newcommand {\vtau}       {v_\tau}
\newcommand {\ahatl}      {\hat{a}_l}
\newcommand {\vhatl}      {\hat{v}_l}
\newcommand {\ahate}      {\hat{a}_e}
\newcommand {\vhate}      {\hat{v}_e}
\newcommand {\ahatmu}     {\hat{a}_\mu}
\newcommand {\vhatmu}     {\hat{v}_\mu}
\newcommand {\ahattau}    {\hat{a}_\tau}
\newcommand {\vhattau}    {\hat{v}_\tau}
\newcommand {\vtildel}    {\tilde{\rm v}_l}
\newcommand {\avsq}       {\ahatl^2\vhatl^2}
\newcommand {\Ahatl}      {\hat{A}_l}
\newcommand {\Vhatl}      {\hat{V}_l}
\newcommand {\Afer}       {A_f}
\newcommand {\Ael}        {A_e}
\newcommand {\Aferb}       {\bar{A_f}}
\newcommand {\Aelb}        {\bar{A_e}}
\newcommand {\AVsq}       {\Ahatl^2\Vhatl^2}
\newcommand {\Iwk}        {I_{3l}}
\newcommand {\Qch}        {|Q_{l}|}
\newcommand {\roots}      {\sqrt{s}}
\newcommand {\mt}         {m_t}
\newcommand {\Rechi}     {{\rm Re} \left\{ \chi (s) \right\}}
\newcommand {\up} {^}
\newcommand {\abscosthe}     {|cos\theta|}
\newcommand {\dsum}     {\Sigma |d_\circ|}
\newcommand {\zsum}     {\Sigma z_\circ}
\newcommand {\sint}      {\mbox{$\sin\theta$}}
\newcommand {\cost}      {\mbox{$\cos\theta$}}
\newcommand {\mcost}      {|\cos\theta|}
\newcommand {\epair}      {\mbox{$e^{+}e^{-}$}}
\newcommand {\mupair}     {\mbox{$\mu^{+}\mu^{-}$}}
\newcommand {\taupair}     {\mbox{$\tau^{+}\tau^{-}$}}
\newcommand {\gamgam}  {\mbox{$e^{+}e^{-}\rightarrow e^{+}e^{-}\mu^{+}\mu^{-}$}}
\newcommand {\mydeg}   {^{\circ}}
\def\lt{\raisebox{0.2ex}{$<$}}
\def\gt{\raisebox{0.2ex}{$>$}}
\def\avthe{$\overline{\theta}$}
\def\avphi{$\overline{\phi}$}
\def\main{$\overline{\rm MAIN}$}
\def\delphi{${\mit\Delta}\phi$}
\newcommand{\fullskip}{\vskip 16cm}
\newcommand{\halfskip}{\vskip  8cm}
\newcommand{\quarskip}{\vskip  6cm}
\newcommand{\abitskip}{\vskip 0.5cm}
\def\Gvis{\Gamma_{\rm vis}}
\def\shad{\sigma_{\rm had}^{\rm pole}}
\def\sll{\sigma_{ll}^{\rm pole}}
\def\see{\sigma_{ee}}
\def\smu{\sigma_{\mu\mu}}
\def\stau{\sigma_{\tau\tau}}
\def\stat{\rm{stat}}
\def\syst{\rm{syst}}

%
\section{Introduction}

In the Supersymmetric extension of the Standard Model, once Supersymmetry is
spontaneously broken the gravitino $\grav$ can acquire a mass by absorbing the
degrees of freedom of the goldstino. This mechanism is analogous to 
the spontaneous breaking of the electro-weak symmetry in the Standard Model, 
where the Z and W bosons
acquire mass by absorbing the goldstone bosons.

A light gravitino 
as predicted by some supersymmetric 
models ~\cite{ref:GMSB} has been searched for at LEP and Tevatron 
experiments~\cite{ref:lepbound,ref:cmsbound}.
Limits on the $\grav$ mass
allow lower limits on 
the supersymmetry-breaking scale $\sqrt{F}$
to be inferred.

Recently it has been pointed out \cite{ref:prz} that an appropriate theory 
must contain also the supersymmetric partner of the goldstino,
called the sgoldstino, which could be massive. The production of 
this particle may be relevant at  present LEP energies if the supersymmetry-breaking
scale and the sgoldstino mass are not too large.
In the minimal R-parity-conserving model, as considered in  \cite{ref:prz},
the effective theory at the weak scale contains two neutral scalar states:
the $S$ which is CP-even, and the $P$ which is CP-odd.
As sgoldstinos have even R parity,  
they are not necessarily produced in pairs and their decay chains 
do not necessarily contain an LSP (Lightest Supersymmetric Particle).
The phenomenology of these two particles is similar. 
The following  formulae and results will be expressed for the $S$
 state but are valid also for the $P$ particle.

At LEP 2, one of the most interesting production channels is
the process  $\eeSg$, which depends on the $S$ mass $m_S$ and on
$\sqrt{F}$:

\begin{equation}
\frac{d \sigma} {dcos\theta} (e^+e^-\rightarrow S\gamma  )
=\frac{\left|\Sigma\right|^2 s}{64 \pi F^2} 
\left( 1- \frac{m_S^2}{s} \right)^3 (1+cos^2\theta),
\label{dsigma}
\end{equation}
where $\theta$ is the scattering angle in the centre-of-mass and

\begin{equation}
\left|\Sigma\right|^2=\frac{e^2 M_{\gamma\gamma}^2}{2s}+
                      \frac{g_Z^2(v_e^2+a_e^2) M_{\gamma Z}^2 s}{2(s-m_Z^2)^2}+
                      \frac{e g_Z v_e M_{\gamma\gamma}M_{\gamma Z}}{s-m_Z^2}
\end{equation}
with $v_e=sin^2 \theta_W -1/4$, $a_e=1/4$ and 
$g_Z=e/(sin \theta_W cos \theta_W)$ .
$M_{\gamma\gamma}$ and  $M_{\gamma Z}$
are related to the diagonal mass terms for the $U(1)_Y$ and $SU(2)_L$ gauginos
$M_1$ and $M_2$:

\begin{equation}
M_{\gamma\gamma}= M_1 cos^2 \theta_W+ M_2 sin^2 \theta_W,~
M_{\gamma Z}= (M_2-M_1) sin \theta_W cos \theta_W.
\end{equation}

The most relevant $S$ decay modes  are $S\rightarrow \gamma \gamma$ and 
$S \rightarrow gg$ with 

\begin{equation}
\Gamma(S\rightarrow \gamma \gamma)=\frac{m_S^3 M_{\gamma\gamma}^2}{32 \pi F^2}
\end{equation}

and 

\begin{equation}
\Gamma(S\rightarrow g g)= \frac{m_S^3 M_3^2}{4 \pi F^2},
\end{equation}
where $M_3$ is the gluino mass. The corresponding branching ratios 
depend on $M_1,~M_2$ and $M_3$,
and the total width is  
$\Gamma\sim \Gamma(S\rightarrow \gamma \gamma)+\Gamma(S\rightarrow g g)$. 
In this letter two sets for these parameters as suggested in \cite{ref:prz} 
are considered and listed in Table \,\ref{tab:param}.
\begin{table}[bth]
\begin{center}
\begin{tabular}{|c|c|c|c|c|c|}
\hline
 &$M_1$&$M_2$& $M_3$ &B.R. $S\rightarrow \gamma \gamma$&B.R. $S\rightarrow g g$   \\      
\hline
1)      & 200   & 300         & 400 & 4$\%$ &96$\%$ \\
\hline
2)      & 350   & 350         & 350 &11$\%$ &89$\%$ \\ 
\hline
\end{tabular}
\caption[]{Two choices for the gaugino 
mass parameters (in GeV/c$^2$) relevant for the sgoldstino production and decay
and the corresponding Branching Ratios of the two considered channels. }
\label{tab:param}
\end{center}
\end{table}

For a large interval of the parameter space 
the total width is  small 
(below a few GeV/c$^2$), except  for the region with small $\sqrt{F}$
where the production cross section is also expected to be very large.


The two decay channels considered produce events with very different 
topologies:
\begin{enumerate}
 \item{ $S\rightarrow \gamma \gamma$ gives rise to
events with three high energy photons, one of which is expected to be 
monochromatic with energy $E_{\gamma}=\frac{s-m_S^2}{2 \sqrt{s} }$ for 
the large fraction of the parameter space
where  $S$ has a negligible
width. Despite  its lower branching ratio
(4 and 11$\%$ for the two sets of Table  \,\ref{tab:param}, respectively),
this final state is worth investigating 
because  the main background source is 
the QED process $\eeggg$,
which is expected to be small if photons in the forward region are discarded. 
}
\item{
 $S\rightarrow  g g$ gives rise to events with one monochromatic 
photon (except for the region with small $\sqrt{F}$ ) and two jets. 
An irreducible background from 
$e^+ e^- \rightarrow q \bar{q} \gamma$ events is associated to this topology.
Therefore the signal must be searched for as an excess of events 
over the  background expectation for every mass hypothesis.
}
\end{enumerate}

This letter  describes the results obtained with the DELPHI detector
at LEP  centre-of-mass energies of 189, 192, 196, 200 and 202 GeV, 
corresponding to a total integrated luminosity of about 380 pb$^{-1}$.

\section{Apparatus}
A detailed description of the DELPHI detector can be found in \cite{ref:delpap}.
The present analysis was mainly based on the measurement of the electromagnetic energy
clusters \cite{ref:nuclph}
 in the barrel electromagnetic calorimeter, the High density
 Projection Chamber (HPC), and in the Forward ElectroMagnetic Calorimeter (FEMC),
as well as on the capability of reconstructing  charged particle tracks using the tracking devices:
the Vertex Detector (VD), the Inner Detector (ID),
the Time Projection Chamber (TPC),
  the Outer Detector (OD) and the forward chambers (FCA and FCB).
The Vertex Detector
\cite{ref:vft} extends its coverage down to
$10.5^{\circ}$ in polar angle $\theta$.
An  electromagnetic calorimeter (STIC) was used 
to measure the luminosity. 

The barrel and the forward electromagnetic energy triggers were based on data 
from the HPC and the FEMC respectively.
The calorimetric trigger efficiency for  $\eeggg$  was
estimated with samples of Bhabha $\ee \rightarrow \ee (\gamma)$  events. 
This was done by counting how
often the electromagnetic  trigger was fired by an electron
 which had been triggered by an independent track trigger.
In events with more than two photons, as well as in events with photons and
charged particle tracks, the trigger efficiency was better than 99$\%$.

\section{Event selection and analysis}
The 1998 data were taken at $\sqrt{s}=188.6$ GeV,
and the 1999 data at
191.6, 195.5, 199.5 and 201.6 GeV.
The integrated luminosities obtained 
requiring 
the HPC, FEMC, TPC and  VD to be operational 
were 
155.1 pb$^{-1}$, 25.1  pb$^{-1}$,
76.2 pb$^{-1}$, 83.1  pb$^{-1}$ and 40.1  pb$^{-1}$ respectively
for the five centre-of-mass energies.
%

Monte-Carlo generated events for the same 
centre-of-mass energies 
 were processed through the full DELPHI
simulation  
$\cite{ref:delpap}$ and the same reconstruction chain as real data. 

\subsection{ $S\rightarrow \gamma \gamma$ channel}



Events were selected as $\gamma \gamma \gamma$ candidates if they 
had:  
\begin{itemize}
\item{ at least two electromagnetic energy clusters  
with $0.219<E / \sqrt{s}<0.713$; 
}
\item{ at least one additional cluster with $E>5 $ GeV and no more than 
two additional clusters, of which  the second one (if present)
had $E<5 $ GeV; }
\item{ the two most energetic electromagnetic clusters in the HPC region 
  $42^{\circ}<\theta<89^{\circ}$ 
     or in the FEMC region
   $25^{\circ}<\theta<32.4^{\circ}$; }
\item{ the third  cluster in the region $42^{\circ}<\theta$ 
or $20^{\circ}<\theta<35^{\circ}$;}
\item{  no hits in two of the three  Vertex Detector layers
within $\pm 2^{\circ}$ in azimuthal angle $\phi$
of the line from the  
mean beam crossing point to any electromagnetic cluster.}
\end{itemize}


Further, two hemispheres were defined by a plane orthogonal to the direction of the most energetic
cluster.  One hemisphere was
required to have no charged particle detected in the barrel region 
of the detector with 
momentum above 1 GeV/c 
extrapolating to within 5 cm of the mean beam crossing point.
The requirement was strengthened, to suppress
the large $e^+e^-$ background further, by demanding that
both hemispheres have no such particle  detected by the TPC 
with $\theta<35^{\circ}$.

The events selected have a three-body final state kinematics
if no significant additional radiation is lost in the  detector (mainly
initial state radiation lost along the beam pipe). A simple way to check
if an event is, within a reasonable approximation, a three-body final state,
is to look at the distribution of the quantity 
$\Delta= \left| \delta_{12}\right| +\left| \delta_{13}\right| +\left| \delta_{23}\right| $
 where $\delta_{ij}$
is the angle between the particle $i$ and $j$  (Fig. \ref{delta}). 
In a three-body final state, 
the particles lie in a plane and therefore $\Delta$ should be $360^{\circ}$. 
If only the events with $\Delta>358^{\circ}$  are accepted, the energies of the 
particles  can be determined with very good precision 
from their measured directions:

\begin{figure}[th]
\begin{center}\mbox{\epsfxsize 10cm\epsfbox{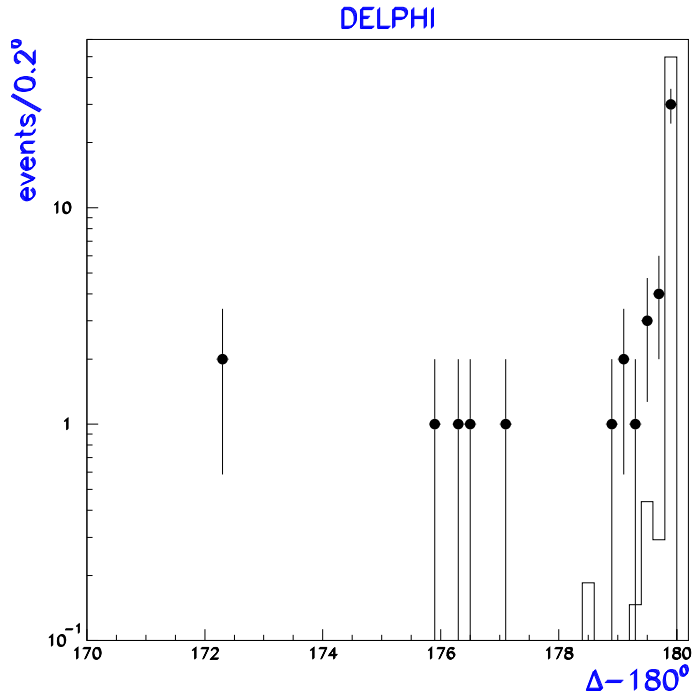}} \end{center}
\caption{$\Delta -180^0$ for the $\gamma \gamma \gamma $ candidates (points)
and the 
 QED $\eeggg$ simulated sample (histogram).
The cuts on  $min(\delta_{12},\delta_{13},\delta_{23})>2^{\circ}$ and on 
 $cos \alpha$ are not applied in this figure.
}
\label{delta}
\end{figure}

\begin{equation}
 E_1= \sqrt{s} \frac{sin \delta_{23}}{\delta};~E_2= \sqrt{s} \frac{sin \delta_{13}}{\delta};~
 E_3= \sqrt{s} \frac{sin \delta_{12}}{\delta}
\label{ecorr}
\end{equation}
with $\delta = sin \delta_{12}+ sin \delta_{13}+  sin \delta_{23}$. The error on the energy 
evaluation was further minimised by requiring 
$min(\delta_{12},\delta_{13},\delta_{23})>2^{\circ}$.

In $S\gamma$ events, the $S$ decay products are isotropically
distributed in the $S$ centre-of-mass system. 
The distribution of $cos \alpha$, where $\alpha$ is the angle between the $S$ 
direction (opposite to the prompt photon) and the direction of one of the
two  $S$ decay products, in the $S$ centre-of-mass system, should 
therefore be flat. 
On the other hand, in the QED 
background $\left| cos \alpha \right|$  peaks at 1.
Therefore, out of the three combinations present in each event, 
only those giving $\left | cos \alpha  \right |<0.9$ 
were accepted.

The numbers of selected events, each giving up to three combinations,
are listed with the expected
background in  
Table \,\ref{tab:evlist}. 
No significant background was found except for the QED process
$\eeggg$.

The  acceptance for an $S\gamma$ signal produced according to
 (\ref{dsigma}) after the described 
polar angle cuts  was $(51\pm 2) \%$. 
The dependence on $m_S$  
from 10 to 190 GeV/c$^2$ was contained within the error quoted. 
The selection efficiency inside
the acceptance region was evaluated by means of the QED background events
generated according to $\cite{ref:kleiss}$,
The efficiency was independent, within the errors, of the photon polar angle.
Its average value  was $(76.6 \pm 2.5) \% $.

\begin{figure}[th]
\begin{center}\mbox{\epsfxsize 9cm\epsfbox{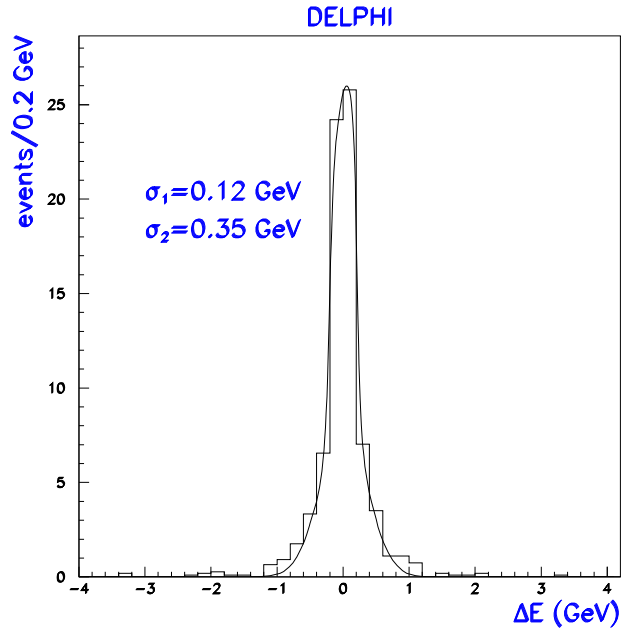}} \end{center}
\caption{
 The  energy resolution $\Delta E$ for the photons of the
$\gamma \gamma \gamma$ candidates in the QED $\eeggg$ simulated sample.
The photon energy was obtained using (\ref{ecorr}).
A fit with two Gaussians gave two resolution components 
 $\sigma_1= 0.12$ GeV and
 $\sigma_2= 0.35$ GeV with approximately equal frequencies. }
\label{resggg}
\end{figure}

The energy resolution obtained from (\ref{ecorr}) was also evaluated
using simulated QED events as shown in 
Fig.\ref{resggg}. It was better than 0.5 $\%$ in the whole photon energy
range:
a fit with two Gaussians  
gave two resolution 
components with  
 $\sigma_1= 0.12$ GeV and
 $\sigma_2= 0.35$ GeV 
with about equal frequencies.

 The second component was introduced to describe 
the tails originating from photons detected near the calorimeter dead regions.

\subsection{ $S\rightarrow  g g$ channel}
This channel is expected to give rise to a final state with one photon 
and two jets. 
An event was selected as a $\gamma g g$ candidate 
if it had:
\begin{itemize}
\item
an electromagnetic energy  cluster identified as a photon with 
$E> 5$ GeV and 
 $\theta>20^{\circ}$;
\item no electromagnetic cluster with $\theta <5^{\circ}$;
\item total multiplicity  greater than 10;
\item charged particle multiplicity  greater than 5;
\item $\sqrt{\sum_{i=1}^n (p_x^2+p_y^2)_i  }> 0.12\times\sqrt{s}$, where $n$ 
is the total multiplicity;
\item the sum of the absolute values of all particle momenta 
along the thrust axis greater than $0.20 \times\sqrt{s}$;
\item either an electromagnetic cluster with $E< 0.45 \times \sqrt{s}$,
or a total multiplicity greater than 16 if the cluster energy is greater than 
$0.45 \times \sqrt{s}$ ;
\item $ |cos(\theta_p)| <0.995$, where $\theta_p$ is the polar angle of
the missing momentum;
\item visible energy  greater than $0.60 \times \sqrt{s}$;
%
\item $|\cos{\alpha}|<0.9$;
\item        $\Delta $  greater than $350 ^{\circ}$.

\end{itemize}

The events were reconstructed forcing all particles but the photon 
into a 2-jet topology using the DURHAM $\cite{durham}$ algorithm.
Events were removed if $ y_{cut}>0.02$
and if the angle between the photon
and the nearest jet was less than $10^{\circ}$.
If the event contained more than one photon candidate, the 
most energetic one was 
considered as the one produced in 
$e^+e^-\rightarrow S\gamma$.
In addition, the jets were required to be incompatible with the 
$b\bar{b}$ hypothesis by requiring 
the combined btag of the events 
to be less than zero $\cite{btag}$.

As in the $\gamma \gamma \gamma$ selection, the events obtained after 
this selection are three-body final state events 
in the absence of additional lost radiation. Therefore all the kinematic constraints
described in the previous subsection were also applied here. 
In this case, however,
as jet directions are less precisely determined than photons directions,
the cut in $\Delta$ was less stringent and
the  resolution for the reconstructed photon energy was  poorer:
a two-Gaussian fit gave $\sigma_1 = 1.2$ GeV ($55 \%$ of the area) and $\sigma_2 = 4.1$ GeV. 

The  polar angle acceptance for an $S\gamma$ signal produced 
according to (\ref{dsigma})
was $(76 \pm2 \%)$ and  almost independent of $m_S$. 
The selection efficiency inside
the acceptance region was evaluated using the $ q \bar{q} \gamma$ background events
generated
with PYTHIA \cite{ref:pythia}, processed through the full 
DELPHI analysis chain and re-weighted according to 
the background and signal photon 
polar angle distributions. It ranged from 20 to 55 $\%$ depending 
on the photon energy.

In addition to the main background from
$ q \bar{q} \gamma$ events, a small (less than 5$\%$) fraction was due 
to four-fermion processes which were generated according 
to  EXCALIBUR \cite{ref:excalibur}.
The numbers of selected events 
and the expected
background are listed in  
Table \,\ref{tab:evlist}.

\begin{table}[bth]
\begin{center}
\begin{tabular}{|c|c|c|c|}
\hline
  channel 			   &$\sqrt{s}$ (GeV)& events       & background\\
\hline
  $S\rightarrow \gamma \gamma$     & 189   	    &11         & $19\pm 2$ \\

  $S\rightarrow \gamma \gamma$     & 192 to 202     &19         & $24^{-2}_{+3}$  \\ 
\hline
  $S\rightarrow g~g$     & 189   	    &771        & $782\pm 24$ \\
  $S\rightarrow g~g$     & 192   	    &113        & $113\pm 3$ \\
  $S\rightarrow g~g$     & 196   	    &339        & $316\pm 5$ \\
  $S\rightarrow g~g$     & 200   	    &342        & $330\pm 6$ \\
  $S\rightarrow g~g$     & 202   	    &169        & $158\pm 3$ \\
\hline
\end{tabular}
\caption[]{Number of selected events for the two decay channels
and expected number of 
background events. 
The background for the  $S\rightarrow \gamma \gamma$  channel is dominated by the QED process $\eeggg$,
                             for the $S\rightarrow g~g$ channel 
                              by the process $e^+ e^- \rightarrow q \bar{q} \gamma$. 
                             The errors include systematic effects (see text).}
\label{tab:evlist}
\end{center}
\end{table}


\section{Results} 
\begin{figure}[th]
\begin{center}\mbox{\epsfxsize 10cm\epsfbox{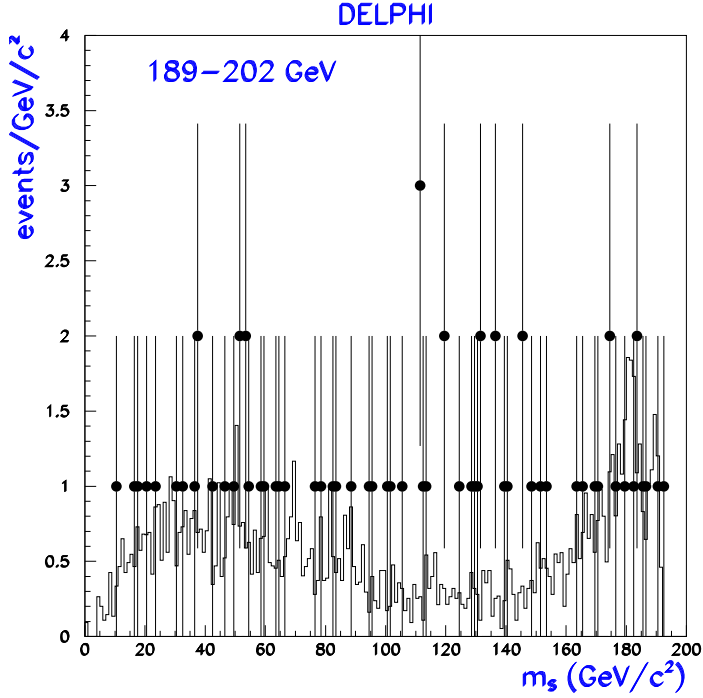}} \end{center}
\caption{
 Photon recoil mass spectrum for the 
$\gamma \gamma \gamma $ candidates (points) and the expected background (histogram).
The average number of entries per event in the data is 2.3. The bin size
takes into account  the experimental mass resolution 
and   the expected signal width. }
\label{msggg}
\end{figure}

\begin{figure}[th]
\begin{center}\mbox{\epsfxsize 9cm\epsfbox{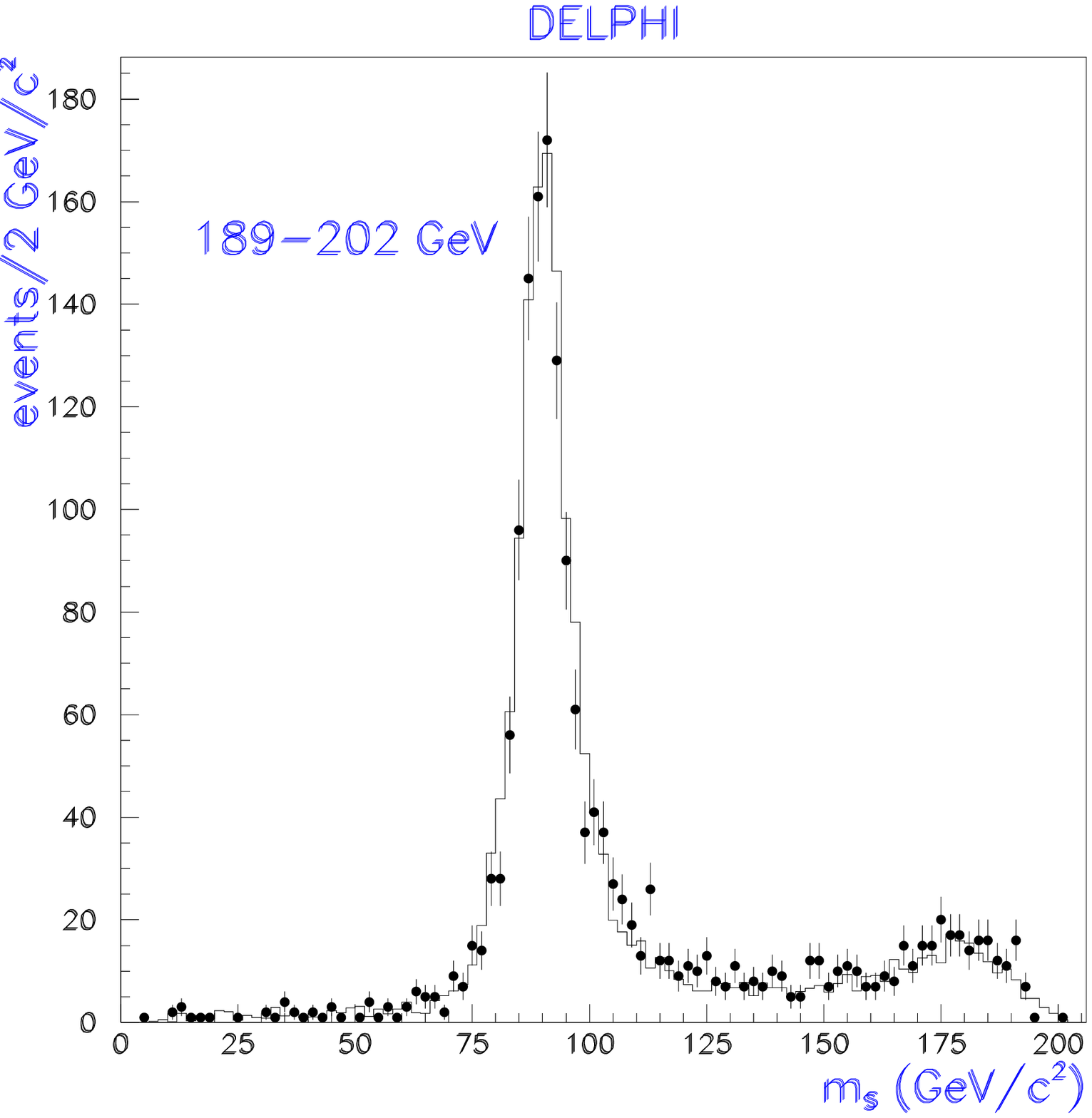}} \end{center}
\caption{
 Photon recoil mass spectrum for the 
$\gamma gg $ candidates (points) and the expected background (histogram).
}
\label{msgglgl}
\end{figure}

The  photon 
recoil mass spectra obtained for  the two decay channels  are shown
in Fig.~\ref{msggg} and Fig.~\ref{msgglgl}. 
The data are superimposed on the expected background distributions. 
In the case of the $S\rightarrow \gamma \gamma$ channel, 
the QED  background generator
included corrections  only to order $\alpha^3$ and therefore 
no additional radiation was simulated. Additional radiation tends 
to give rise to a tail of events  having low values 
of $\Delta$  (Fig. \ref{delta}).   
These events were removed only from the selected sample of real data,
and therefore
a corresponding normalisation correction of 
$(-13^{+4}_{-7}) \%$ 
was applied to the simulated sample. This correction was
the dominant contribution  to 
the systematic uncertainty for the $S\rightarrow \gamma \gamma$ channel. 

In the case of the $S\rightarrow g g$ channel, the systematic error was due
to the Monte-Carlo statistics and  to the uncertainty on the luminosity
determination, which was $0.56\%$ for the 
1998 data and $1.0 \%$ for the 1999 data. The  
 $e^+ e^- \rightarrow q \bar{q} \gamma$ background for the 189 GeV data
was generated with PYTHIA version  5.722, which did not
accurately   reproduce the 
angular distribution of the radiative photon. Therefore the 
Monte-Carlo events at that energy were corrected on the basis 
of the ratio between events generated at higher energies according to 
PYTHIA version  5.722 and  PYTHIA version 6.125. The systematic error on the
number of expected events at 189 GeV includes the uncertainty in this correction. 

No excess of events and no clear evidence of 
anomalous production of events with monochromatic photons is observed in 
either channel. 
Therefore   a limit on the cross section of the new physics reaction
contributing to the two topologies was set.

\begin{figure}[th]
\begin{center}\mbox{\epsfxsize 13cm\epsfbox{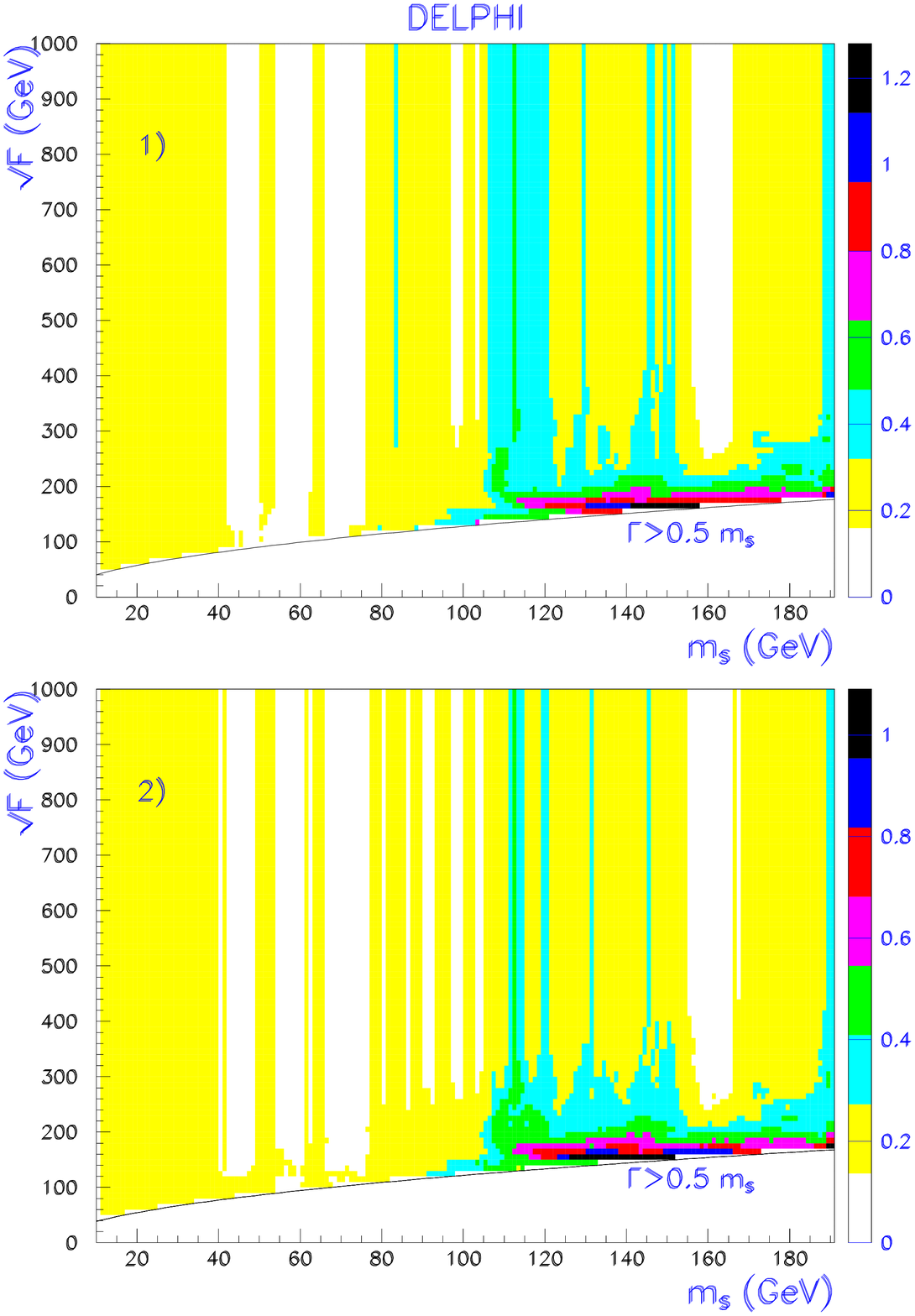}} \end{center}
\caption{
 Cross section (pb) upper limit at the 95$\%$ Confidence Level 
   as a function of $m_{S}$ and $\sqrt{F}$ for the two
 sets of parameters of Tab. \ref{tab:param}.  
}
\label{cslim}
\end{figure}

\begin{figure}[th]
\begin{center}\mbox{\epsfxsize 13cm\epsfbox{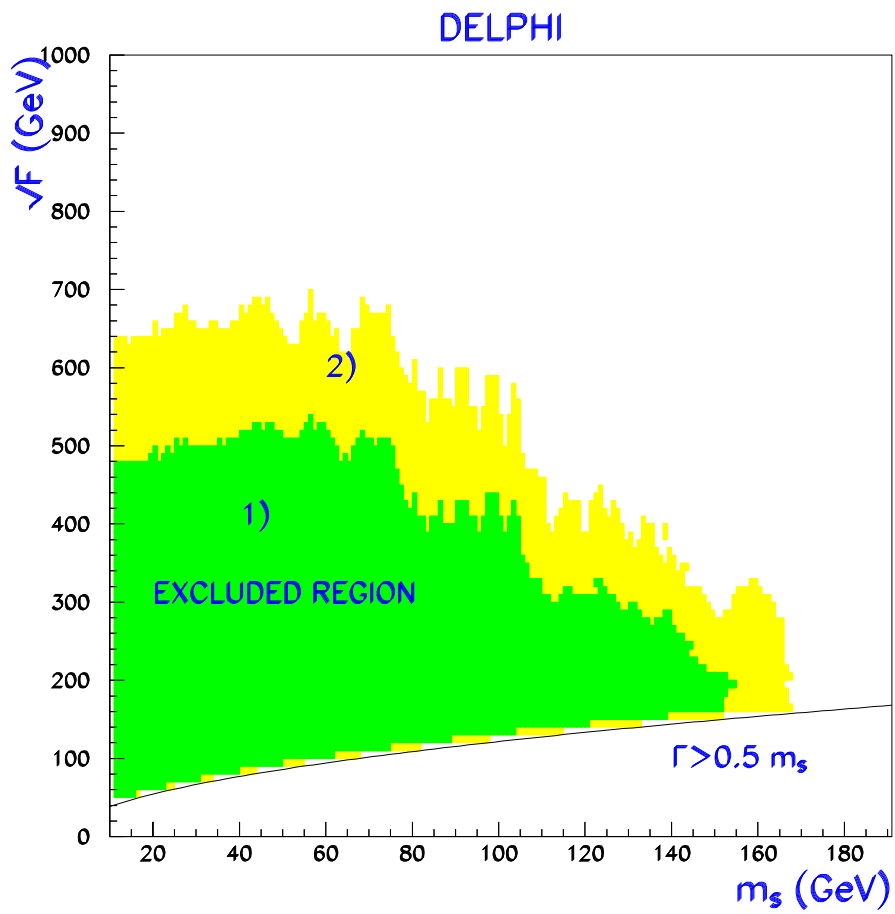}} \end{center}
\caption{
 Exclusion region at the 95$\%$ Confidence Level in the $m_{S}$,  $\sqrt{F}$ plane
 for the two
 sets of parameters of Tab. \ref{tab:param}.  
}
\label{excl}
\end{figure}

\begin{figure}[th]
\begin{center}\mbox{\epsfxsize 13cm\epsfbox{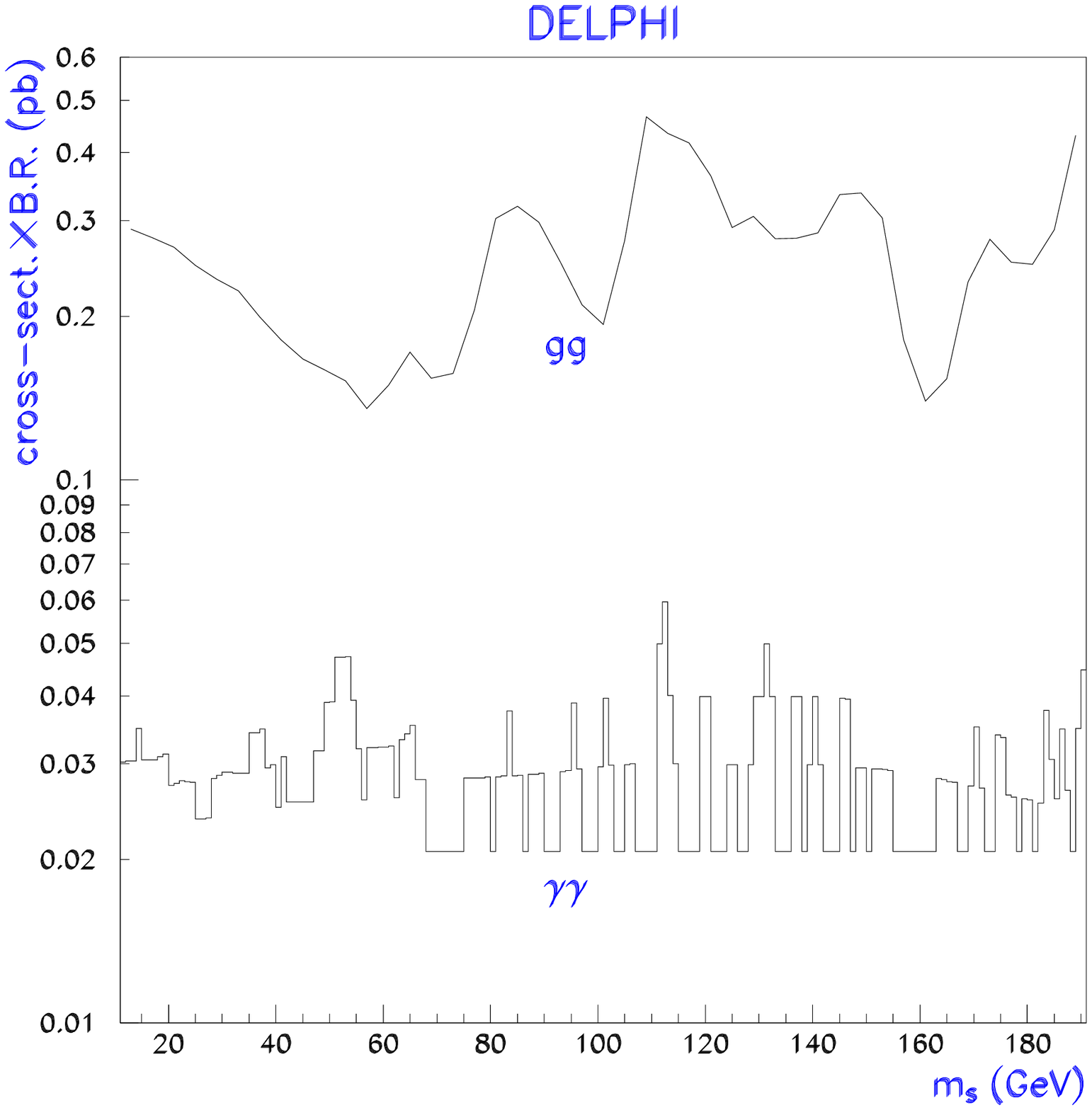}} \end{center}
\caption{
Cross section times branching ratio limits at the 95$\%$ Confidence Level
 for the two decay channels investigated.
 They are obtained for
 $\sqrt{F} \geq 500 $ GeV, corresponding to the region where the expected signal width
 is dominated by the experimental resolution.  
 The bin size
 was chosen to match the experimental mass resolution. }
\label{chann}
\end{figure}

The number of detected events, the background rate 
and the detection efficiency depend 
on the $S$ mass hypothesis considered. In addition, when the expected total
width for a given $m_S$ value is comparable with the 
experimental resolution or larger, the data were compared with the background events
in a region corresponding to
$80 \%$ of the signal area. 
As a consequence, the limit on the signal cross section depends 
on both $m_S$ and $\sqrt{F}$. 
To take into account  
the different sensitivities of the two analysed
channels, the likelihood ratio method 
was used
\cite{ref:lrat}. 
Since the expected $S$  branching ratio and total width depend on the mass 
parameters, as explained above, the 
$95\%$ Confidence Level  cross section limit 
was computed as
a function of $m_S$ and $\sqrt{F}$ for the two sets of parameters listed 
in Table\,\ref{tab:param}. The result is shown in Fig. \ref{cslim}. 
By comparing the experimental limits with the production cross section 
computed from (\ref{dsigma}), it is possible to determine a  $95 \%$ Confidence Level
excluded region of the parameter space. This is shown in Fig. \ref{excl}. 
As explained in
\cite{ref:prz},
to keep the particle interpretation the total width $\Gamma$ must be 
much smaller than $m_S$ and therefore the region with $\Gamma > 0.5 \times m_S$
was not considered. 
The 
$95\%$ Confidence Level limits on the cross section times branching ratio for the two
decay channels are given in Fig. \ref{chann}. They are obtained for 
$\sqrt{F} \geq 500 $ GeV, corresponding to the region where the expected signal width
is independent of $\sqrt{F}$ as it is dominated by the experimental resolution.

\section{Conclusions}
The first search for the production of $S  \gamma$
$(P \gamma)$ where $S~(P)$ is a CP-even (CP-odd) state of the sgoldstino, the goldstino supersymmetric partner,
was made using the 
data collected by DELPHI at LEP in 1998 and 1999 at 
centre-of-mass energies from 189 to 202~GeV for a total integrated 
luminosity of about 380 pb$^{-1}$. 
The $\gamma \gamma \gamma $ and $ \gamma g g $ final states 
expected from  $S(P)\rightarrow \gamma \gamma$ and 
$S(P)\rightarrow g g$ production and decay respectively, were studied. 
No evidence of a signal was found in either channel.
Upper limits on $S \gamma$  ($P \gamma$) production
in the ($m_S$($m_P$), $ \sqrt{F} $) plane
were derived. 

\subsection*{Acknowledgements}
\vskip 3 mm
We want to thank  F. Zwirner for useful explanations of the
theoretical framework and 
for suggestions for possible experimental investigations.
We acknowledge in particular the support of \\
Austrian Federal Ministry of Science and Traffics, GZ 616.364/2-III/2a/98, \\
FNRS--FWO, Flanders Institute to encourage scientific and 
technological research in the industry (IWT), Belgium,  \\
FINEP, CNPq, CAPES, FUJB and FAPERJ, Brazil, \\
Czech Ministry of Industry and Trade, GA CR 202/96/0450 and GA AVCR A1010521,\\
Danish Natural Research Council, \\
Commission of the European Communities (DG XII), \\
Direction des Sciences de la Mati$\grave{\mbox{\rm e}}$re, CEA, France, \\
Bundesministerium f$\ddot{\mbox{\rm u}}$r Bildung, Wissenschaft, Forschung 
und Technologie, Germany,\\
General Secretariat for Research and Technology, Greece, \\
National Science Foundation (NWO) and Foundation for Research on Matter (FOM),
The Netherlands, \\
Norwegian Research Council,  \\
State Committee for Scientific Research, Poland, 2P03B06015, 2P03B11116 and
SPUB/P03/DZ3/99, \\
JNICT--Junta Nacional de Investiga\c{c}\~{a}o Cient\'{\i}fica 
e Tecnol$\acute{\mbox{\rm o}}$gica, Portugal, \\
Vedecka grantova agentura MS SR, Slovakia, Nr. 95/5195/134, \\
Ministry of Science and Technology of the Republic of Slovenia, \\
CICYT, Spain, AEN96--1661 and AEN96-1681,  \\
The Swedish Natural Science Research Council,      \\
Particle Physics and Astronomy Research Council, UK, \\
Department of Energy, USA, DE--FG02--94ER40817. \\

We are also greatly indebted to our technical
collaborators and to the funding agencies for their
support in building and operating the DELPHI detector, and to the members
of  the CERN-SL Division for the excellent performance of the LEP collider.

\end{document}